\documentclass[conference,final,10pt,a4paper]{IEEEtran}

\usepackage[boxed,noline]{algorithm2e}
\usepackage{amsmath}
\usepackage{amssymb}
\usepackage{balance}
\usepackage{cite}
\usepackage{xcolor}
\usepackage[utf8]{inputenc}
\usepackage[T1]{fontenc}
\usepackage{graphicx}
\usepackage{multirow}
\usepackage{textcomp}
\usepackage{marvosym}
\usepackage{wasysym}
\usepackage[colorlinks,bookmarksopen,bookmarksnumbered,citecolor=red,urlcolor=red]{hyperref}
\graphicspath{{figures/}}	

\begin{document}

\title{\LARGE Stacked-VLAN-Based Modeling of Hybrid ISP Traffic Control Schemes
  and Service Plans Exploiting Excess Bandwidth in Shared Access Networks}

\author{%
  \IEEEauthorblockN{Kyeong Soo Kim}%
  \IEEEauthorblockA{%
    Department of Electrical and Electronic Engineering\\%
    Xi'an Jiaotong-Liverpool University\\%
    Suzhou, 215123, P. R. China\\%
    Email: Kyeongsoo.Kim@xjtlu.edu.cn%
  }%
}%

\hypersetup{%
  pdfkeywords={Access, service plan, network pricing, resource allocation,
    Internet service provider (ISP), traffic shaping, quality of service
    (QoS)},%
  pdfsubject={OMNeT++ 2016 paper on fully shared access networks},%
  pdfcreator={PDFLaTeX}%
}

\maketitle

\begin{abstract}
  The current practice of shaping subscriber traffic using a token bucket filter
  by Internet service providers may result in a severe waste of network
  resources in shared access networks; except for a short period of time
  proportional to the size of a token bucket, it cannot allocate excess
  bandwidth among active subscribers even when there are only a few active
  subscribers. To better utilize the network resources in shared access
  networks, therefore, we recently proposed and analyzed the performance of
  access traffic control schemes, which can allocate excess bandwidth among
  active subscribers proportional to their token generation rates. Also, to
  exploit the excess bandwidth allocation enabled by the proposed traffic
  control schemes, we have been studying flexible yet practical service plans
  under a hybrid traffic control architecture, which are attractive to both an
  Internet service provider and its subscribers in terms of revenue and quality
  of service. In this paper we report the current status of our modeling of the
  hybrid traffic control schemes and service plans with OMNeT++/INET-HNRL based
  on IEEE standard 802.1Q stacked VLANs.
\end{abstract}


\pagestyle{empty} 
\thispagestyle{empty}

\section{Introduction}
\label{sec:introduction}
\begin{figure}[!htb]
  \begin{center}
    \includegraphics[angle=-90,width=\linewidth,trim=15 0 0
    0]{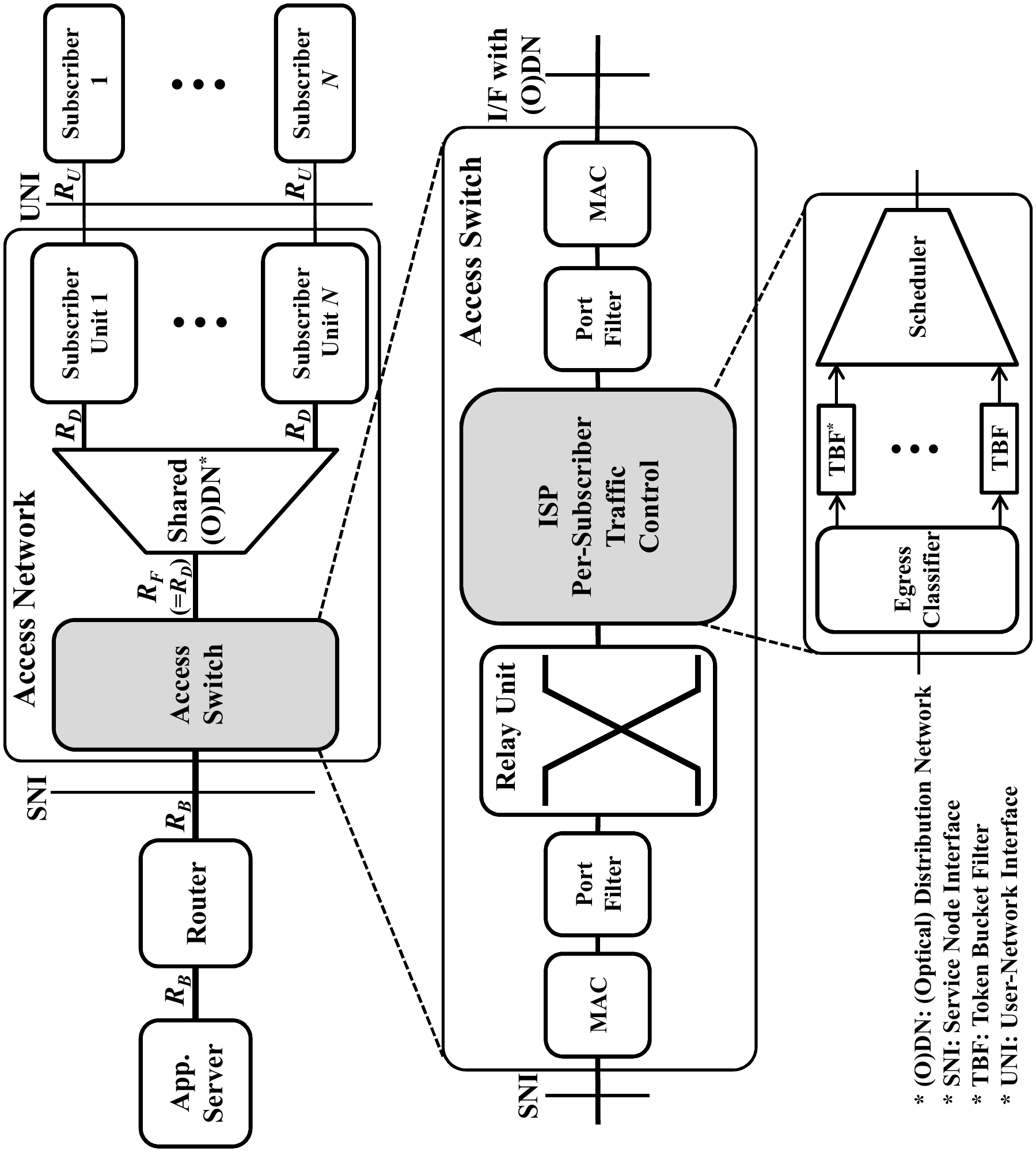}
  \end{center}
  \caption{Overview of current practice of ISP traffic control in shared access
    (shown for downstream traffic only) \cite{Kim:14-5}.}
  \label{fig:isp_traffic_control}
\end{figure}

The resource sharing in shared access networks --- like cable Internet based on
hybrid fiber-coaxial (HFC) networks or passive optical networks (PONs) --- is a
key to achieving lower infrastructure cost and higher energy efficiency. The
full sharing of the bandwidth available among subscribers in a shared access
network, however, is hindered by the current practice of traffic control by
Internet service providers (ISPs), which is illustrated in
Fig.~\ref{fig:isp_traffic_control};
due to the arrangement of traffic shapers (i.e., token bucket filters (TBFs))
and a scheduler in the access switch, the capability of allocating available
bandwidth by the scheduler is limited to the \emph{traffic already shaped} per
service contracts with subscribers \cite{bauer11:_power,Kim:13-3}.

Even though the allocation of excess bandwidth in a shared link has been
discussed in the general context of quality of service (QoS) control (e.g.,
\cite{HTB}), it is recently when the issue was studied in the specific context
of ISP traffic control in shared access \cite{Kim:14-1,Kim:15-3}.
Based on the ISP traffic control schemes proposed in \cite{Kim:14-1} and
\cite{Kim:15-3}, we have been studying the design of flexible yet practical ISP
service plans exploiting the excess bandwidth allocation in shared access
networks
under a hybrid ISP traffic control architecture in order to gradually introduce
the excess bandwidth allocation while providing backward compatibility with the
existing traffic control infrastructure \cite{Kim:14-5}.
To the best of our knowledge, our work in \cite{Kim:14-1,Kim:14-5,Kim:15-3} is
the first effort to study the issue of enabling excess bandwidth allocation
among the subscribers, together with its business aspect, in the context of ISP
traffic control in shared access.

In this paper, we report the current status of our modeling of the hybrid ISP
traffic control schemes and service plans exploiting excess bandwidth in shared
access networks with OMNeT++ \cite{Varga:01} and INET-HNRL\footnote{A fork of
  INET framework (rev. INET-20111118) \cite{INET} and available at
  \url{http://github.com/kyeongsoo/inet-hnrl}, which requires OMNeT++ version
  4.6 and later.} based on the stacked virtual local area networks (VLANs) of
IEEE standard 802.1Q \cite{IEEE:802.1q-2014}.

\section{Review of Hybrid ISP Traffic Control for Shared Access}
\label{sec:traff-contr-schem}
In this section, we briefly review the hybrid ISP traffic control schemes and
service plans for shared access that we proposed in \cite{Kim:14-5}.

\begin{figure}[!t]
  \centering
  \includegraphics[angle=-90,width=.9\linewidth,trim=0 0 0
  0]{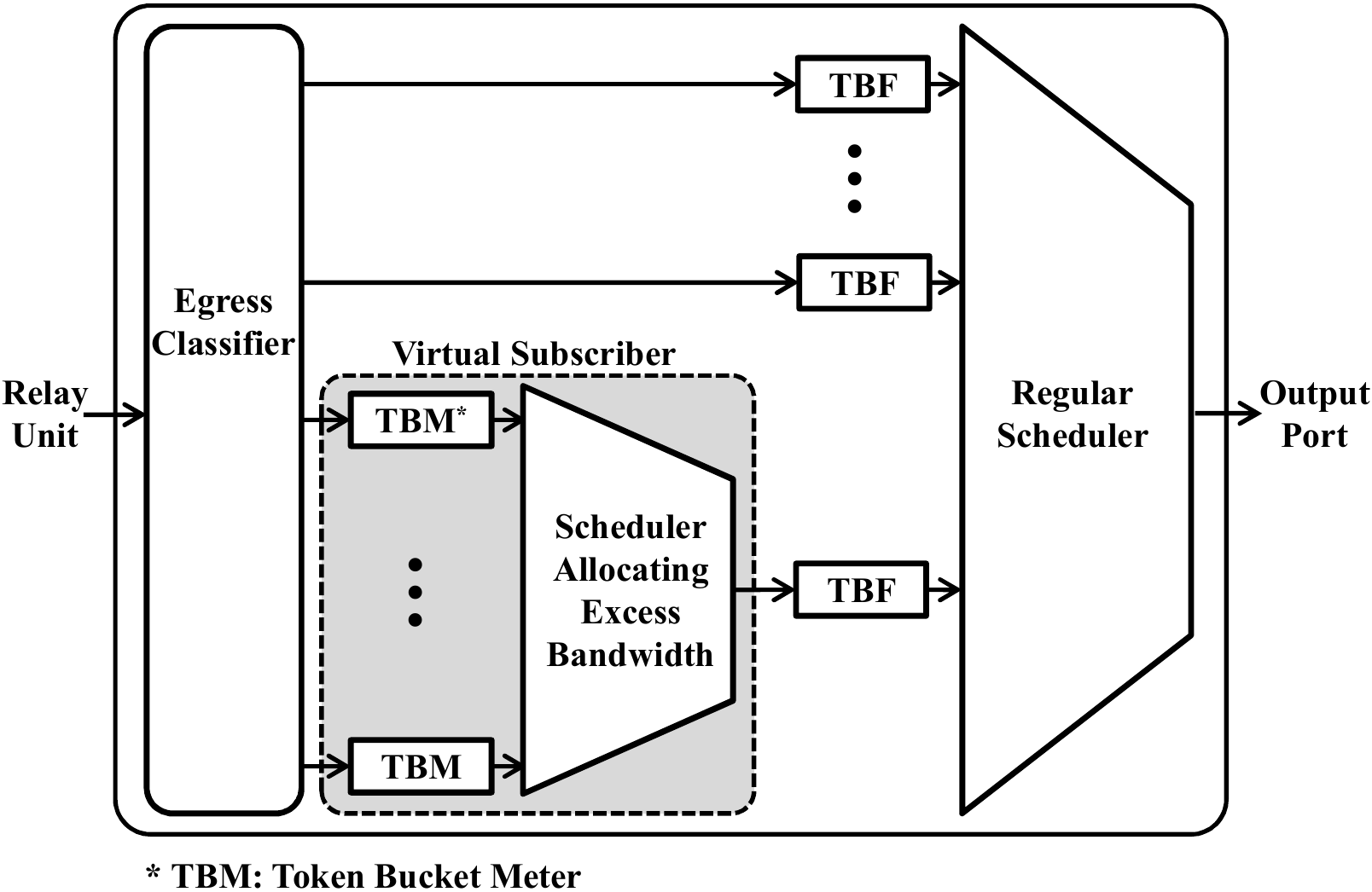}
  \caption{Hybrid ISP traffic control for a flexible service plan exploiting
    excess bandwidth allocation \cite{Kim:14-5}.}
  \label{fig:hybrid_traffic_control}
\end{figure}

Fig.~\ref{fig:hybrid_traffic_control} shows the proposed architecture for hybrid
ISP traffic control, where there coexist subscribers for the current flat-rate
service plan and those for a new service plan fully sharing the bandwidth among
them. For backward compatibility with the existing traffic control and pricing
schemes, the new service plan subscribers are grouped together and treated as
one \emph{virtual} subscriber under the flat-rate service plan; at the same
time, the traffic from each subscriber of the new service plan is individually
controlled by an ISP traffic control scheme enabling excess bandwidth allocation
within the group.
The migration toward fully-shared access will be completed when all the
subscribers of the flat-rate service plan move to the new service plan
exploiting excess bandwidth allocation.

Note that, for the new service plan to be acceptable, it is desirable that there
should be no disadvantage in adopting the new service plan for both ISP and its
subscribers compared to the existing flat-rate service plan. In this regard, we
can derive requirements for the new service plan to meet in terms of parameters
for existing flat-rate service plans, including monthly price, token generation
rate, and token bucket size. Interested readers are referred to \cite{Kim:14-5}
for details.

\section{Modeling of Hybrid ISP Traffic Control Schemes and Service Plans Based
  on Stacked-VLANs}
\label{sec:modeling-hybrid-isp}
As discussed in \cite{Kim:14-2}, we have already implemented models of the
shared access network shown in Fig.~\ref{fig:isp_traffic_control} based on VLAN
as part of INET-HNRL, because we want abstract models that can provide features
common to specific systems (e.g., cable Internet and Ethernet PON (EPON)), while
being practical enough to be compatible with other components and systems of the
whole network.
In the VLAN-based shared access models, we use a VLAN identifier (VID) to
identify each subscriber, which is similar to the service identifier (SID) in
cable Internet and the logical link identifier (LLID) in EPON.

For the implementation of models for the hybrid ISP traffic control shown in
Fig.~\ref{fig:hybrid_traffic_control}, we can think of two distinct approaches,
i.e., an integrated approach where we implement the whole scheduling as one
system (e.g., based on the hierarchical token bucket (HTB) scheduler \cite{HTB})
and a modular approach where we integrate separate schedulers (e.g., a scheduler
based on TBF shaping and a DRR-based scheduler enabling excess bandwidth
allocation \cite{Kim:15-3}) into one. Considering the ease of the management of
two separate groups of subscribers and the upgradability of the component
scheduler allocating excess bandwidth independently of the traditional one based
on TBFs, we have chosen a modular approach and again based our implementation on
VLAN.

\begin{figure}[!tb]
  \centering
  \includegraphics[angle=-90,width=\linewidth]{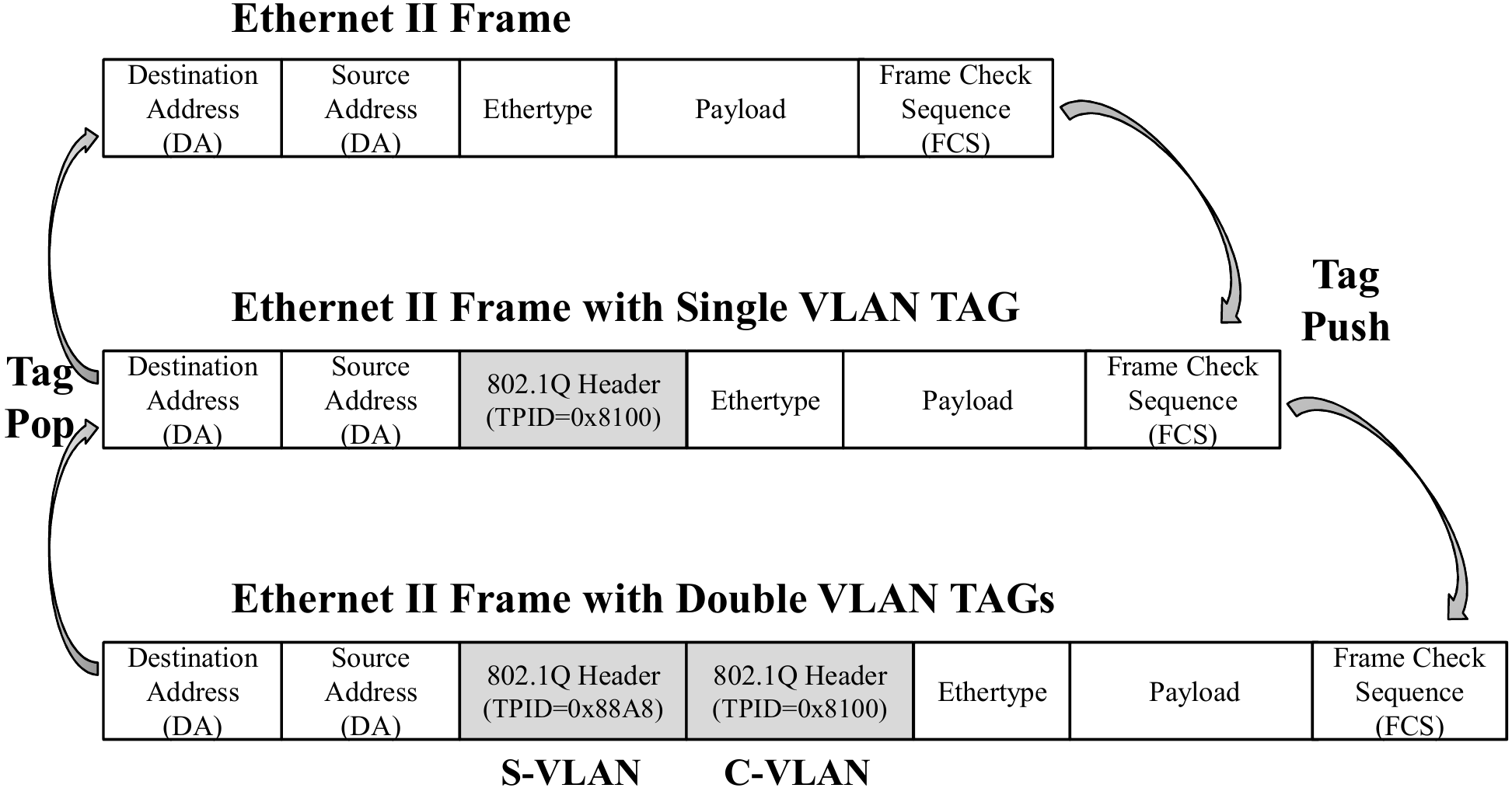}
  \caption{Frame formats for VLAN stacking.}
  \label{fig:frame_formats}
\end{figure}

Unlike existing models based on a single VLAN tag per frame, we need two
different ways of identifying frames from the subscribers for the new hybrid
traffic control scheme and service plan: As for the existing TBF-based traffic
control scheme, the whole frames from those subscribers need to be identified
and treated as a group (i.e., one virtual subscriber) for traffic shaping and
scheduling; as for the new excess-bandwidth-allocating traffic control scheme,
on the other hand, the frames from each subscriber need to be identified and
treated as a separate flow. Fortunately, this requirement of hierarchical
identification of Ethernet frames under the new hybrid traffic control scheme
can be met by the technique of \textit{stacked VLANs} (also called
\textit{provider bridging} and \textit{Q-in-Q}), which is now part of IEEE
standard 802.1Q \cite{IEEE:802.1q-2014}. The change of Ethernet frame formats
related with the VLAN stacking and two tag operations are shown in
Fig.~\ref{fig:frame_formats}.
Note that the tag protocol identifier (TPID) of the second service VLAN (S-VLAN)
tag is set to a value of 0x88A8, different from the value of 0x8100 for the
first customer VLAN (C-VLAN) tag.

\begin{figure*}[!tb]
  \begin{center}
    \includegraphics[angle=-90,width=\linewidth]{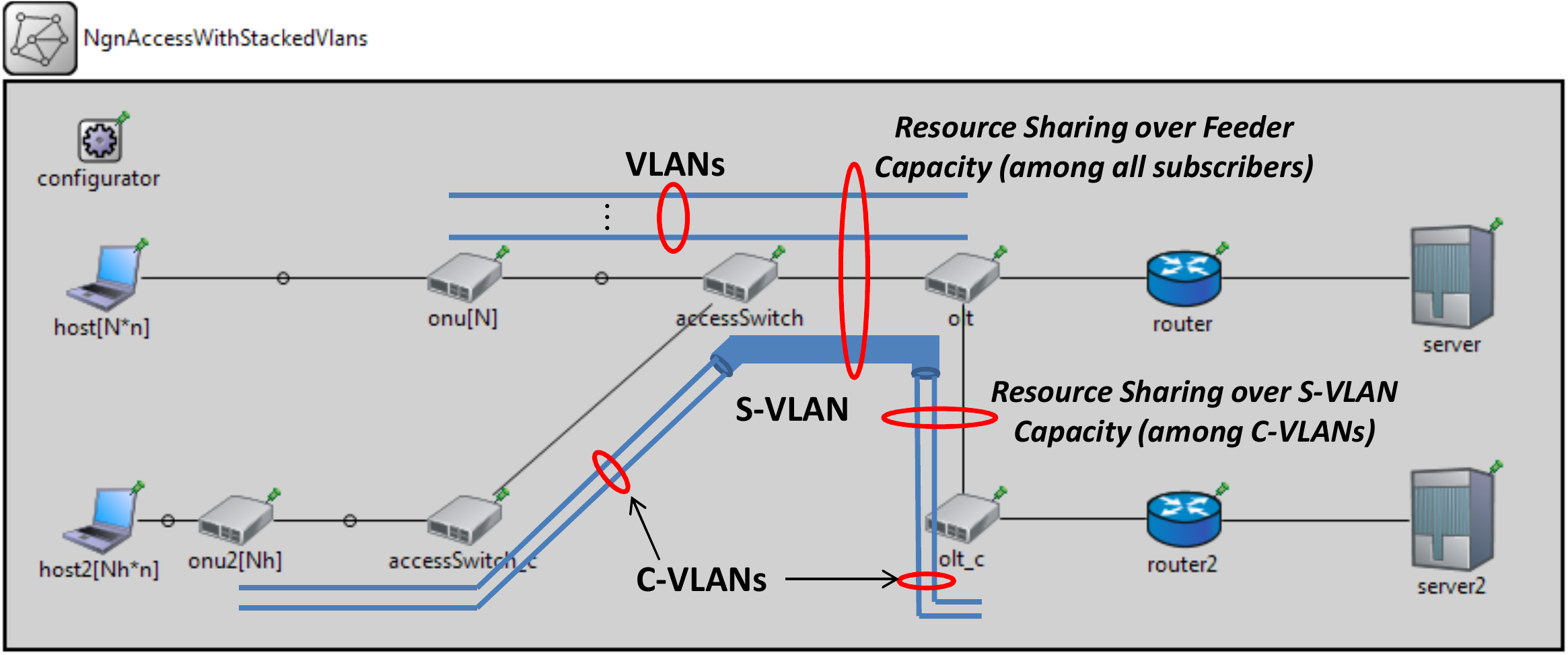}
  \end{center}
  \caption{Stacked-VLAN-based modeling of an access network with hybrid ISP
    traffic control.}
  \label{fig:access_model}
\end{figure*}

\begin{figure}
  \begin{center}
    \includegraphics[width=.55\linewidth]{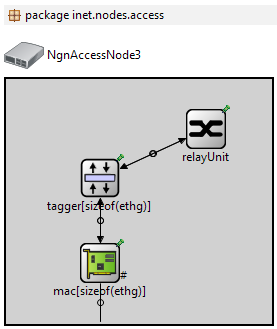}
  \end{center}
  \caption{Ethernet switch module (\textit{NgnAccessNode3}) with stacked-VLAN
    capabilities (for ONUs, OLTs, and access switches shown in
    Fig.~\ref{fig:access_model}).}
  \label{fig:vlan_switch}
\end{figure}

\begin{figure}
  \begin{center}
    \includegraphics[width=.5\linewidth]{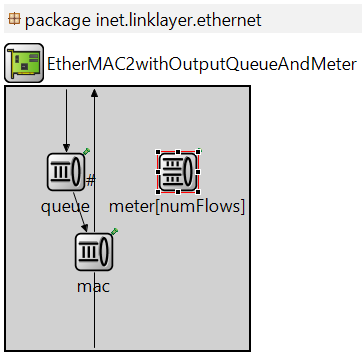}
  \end{center}
  \caption{Ethernet MAC (\textit{EtherMAC2}) module with a queue and a traffic
    meter for modeling of hybrid ISP traffic control.}
  \label{fig:ethermac2}
\end{figure}

Fig.~\ref{fig:access_model} shows stacked-VLAN-based modeling of a shared access
network with hybrid ISP traffic control, while Figs.~\ref{fig:vlan_switch} and
\ref{fig:ethermac2} show the Ethernet switch module for ONUs, OLTs, and access
switches, and the Ethernet MAC module implementing hybrid ISP traffic control,
respectively; as for the traffic control schemes enabling excess bandwidth
allocation, there are implemented two queue types, i.e., \textit{CSFQVLANQueue5}
for the algorithm based on core-stateless fair queueing (CSFQ) \cite{Kim:14-1}
and \textit{DRRVLANQueue3} for the algorithm based on deficit round-robin (DRR)
\cite{Kim:15-3}.

First, the ``olt\_c'' access switch carries out individual traffic control based
on the customer VID (C-VID) of a frame with a single C-VLAN tag, which is
assigned to each subscriber, and sends resulting frames to the second access
switch node ``olt''. At the ``olt'', the C-VLAN frames are grouped together with
the second S-VLAN tag (i.e., VLAN stacking) and go through another traffic
control together with frames from other subscribers with normal (i.e.,
unstacked) VLAN tags. In this way, traffic for the subscribers of the new hybrid
traffic control scheme and service plan goes through two stages of traffic
control, i.e., one at the ``olt\_c'' exploiting excess bandwidth allocation and
the other at the ``olt'' based on traditional TBF-based traffic shaping.

In implementing models of the hybrid traffic control in shared access based on
stacked VLANs, we tried to meet the following major requirements:
\begin{itemize}
\item \textit{Backward compatibility} with the existing VLAN implementations in
  INET-HNRL, including
  \begin{itemize}
  \item \textit{EthernetFrameWithVLAN} message format
  \item \textit{MACRelayUnitNPWithVLAN} and \textit{VLANTagger} modules
  \end{itemize}
\item \textit{Expandability} to stack more than two VLAN tags
\end{itemize}

\begin{figure*}[!tb]
\centering
\includegraphics[width=1.0\textwidth]{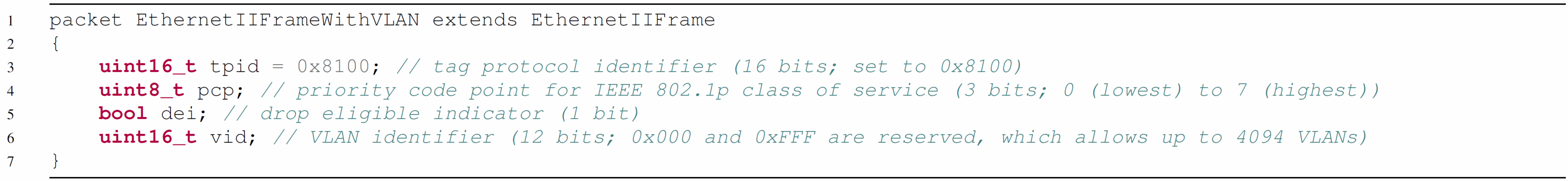}%

    {\scriptsize (a)}\\[0.75em]

%
%
    %

\includegraphics[width=1.0\textwidth]{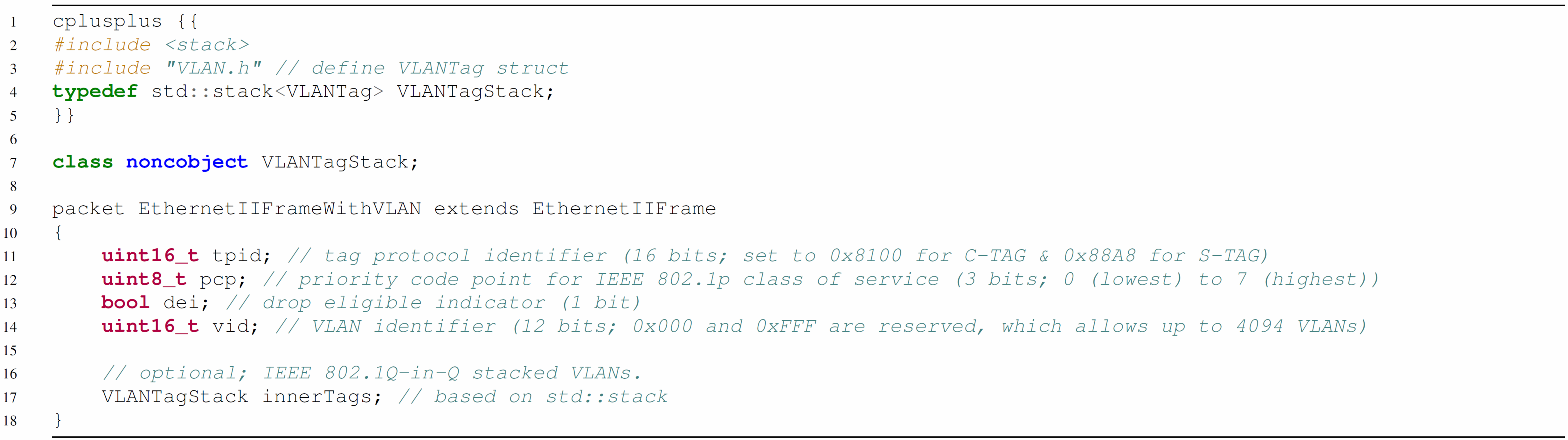}%

    {\scriptsize (b)}

    \caption{Message definitions of Ethernet II frame with VLAN support: (a)
    Without and (b) with VLAN stacking.}
  \label{fig:message_defs} 
\end{figure*}

Consider the original definition of \textit{EthernetFrameWithVLAN} message shown
in Fig.~\ref{fig:message_defs}~(a). Because the \textit{MACRelayUnitNPWithVLAN}
switching module is based on the \textit{vid} field of the
\textit{EthernetFrameWithVLAN} message, which is directly accessible by the
\textit{getVid()} member function, we had to keep these fields in the new
definition of \textit{EthernetFrameWithVLAN} message. For stacking of VLAN tags,
on the other hand, we need to introduce \textit{innerTags} field based on the
\textit{stack} C++ type, which is shown in Fig.~\ref{fig:message_defs}~(b) and
ignored by the existing modules based on the original definition of
\textit{EthernetFrameWithVLAN} message, including
\textit{MACRelayUnitNPWithVLAN} module. In this way, we can meet both the
requirements.


Note that in the current implementation of stacked VLANs, broadcasting is not
allowed across the hierarchies of stacked VLANs. In the shared access network
model shown in Fig.~\ref{fig:access_model}, for example, broadcasting is
possible among normal VLANs or C-VLANs within the same S-VLAN. Broadcasting over
the hierarchies of stacked VLANs requires the modification of the learning
mechanism implemented in the current \textit{MACRelayUnitNPWithVLAN} module.

\section{Summary}
\label{sec:summary}
In this paper we discuss the issues in current practice of ISP traffic shaping
and related flat-rate service plans in shared access networks and review
alternative service plans based on new hybrid ISP traffic control schemes
exploiting excess bandwidth. We also report the current status of our modeling
of the hybrid ISP traffic control schemes and service plans with
OMNeT++/INET-HNRL based on stacked VLANs.

In implementing models of the hybrid traffic control in shared access based on
stacked VLANs, we maintain backward compatibility with the existing modules for
Ethernet switching and VLAN tagging and yet enable the support of stacking of
multiple VLAN tags by clever modification of the message definition for Ethernet
frame with VLAN tags.

\section*{Acknowledgment}
This work was supported by Xi'an Jiaotong-Liverpool University Research
Development Fund (RDF) under grant reference number RDF-14-01-25.



\end{document}